\PassOptionsToPackage{unicode}{hyperref}
\PassOptionsToPackage{hyphens}{url}
\PassOptionsToPackage{dvipsnames,svgnames,x11names}{xcolor}
\documentclass[
]{article}
\usepackage{amsmath,amssymb}
\usepackage{lmodern}
\usepackage{iftex}
\ifPDFTeX
  \usepackage[T1]{fontenc}
  \usepackage[utf8]{inputenc}
  \usepackage{textcomp} 
\else 
  \usepackage{unicode-math}
  \defaultfontfeatures{Scale=MatchLowercase}
  \defaultfontfeatures[\rmfamily]{Ligatures=TeX,Scale=1}
\fi
\IfFileExists{upquote.sty}{\usepackage{upquote}}{}
\IfFileExists{microtype.sty}{
  \usepackage[]{microtype}
  \UseMicrotypeSet[protrusion]{basicmath} 
}{}
\makeatletter
\@ifundefined{KOMAClassName}{
  \IfFileExists{parskip.sty}{%
    \usepackage{parskip}
  }{
    \setlength{\parindent}{0pt}
    \setlength{\parskip}{6pt plus 2pt minus 1pt}}
}{
  \KOMAoptions{parskip=half}}
\makeatother
\usepackage{xcolor}
\usepackage{graphicx}
\makeatletter
\def\maxwidth{\ifdim\Gin@nat@width>\linewidth\linewidth\else\Gin@nat@width\fi}
\def\maxheight{\ifdim\Gin@nat@height>\textheight\textheight\else\Gin@nat@height\fi}
\makeatother
\setkeys{Gin}{width=\maxwidth,height=\maxheight,keepaspectratio}
\makeatletter
\def\fps@figure{htbp}
\makeatother
\providecommand{\tightlist}{%
  \setlength{\itemsep}{0pt}\setlength{\parskip}{0pt}}
\newlength{\cslhangindent}
\newlength{\csllabelwidth}
\newlength{\cslentryspacingunit} 
\newenvironment{CSLReferences}[2] 
 {
  \setlength{\parindent}{0pt}
  \ifodd #1
  \let\oldpar\par
  \def\par{\hangindent=\cslhangindent\oldpar}
  \fi
  \setlength{\parskip}{#2\cslentryspacingunit}
 }%
 {}
\usepackage{calc}

\ifLuaTeX
\usepackage[bidi=basic]{babel}
\else
\usepackage[bidi=default]{babel}
\fi
\babelprovide[main,import]{american}

\def\languageshorthands#1{}
\ifLuaTeX
  \usepackage{selnolig}  
\fi
\IfFileExists{bookmark.sty}{\usepackage{bookmark}}{\usepackage{hyperref}}
\IfFileExists{xurl.sty}{\usepackage{xurl}}{} 
\hypersetup{
  pdftitle={doped: Python toolkit for robust and repeatable charged
defect supercell calculations},
  pdfauthor={Seán R. Kavanagh, Alexander G. Squires, Adair Nicolson,
Irea Mosquera-Lois, Alex M. Ganose, Bonan Zhu, Katarina Brlec, Aron
Walsh, David O. Scanlon},
  pdflang={en-US},
  colorlinks=true,
  linkcolor={Maroon},
  filecolor={Maroon},
  citecolor={Blue},
  urlcolor={Blue},
  pdfcreator={LaTeX via pandoc}}

\title{doped: Python toolkit for robust and repeatable charged defect
supercell calculations}

\usepackage[affil-it]{authblk}
\usepackage{orcidlink}
\author[1,2%
  \ensuremath\mathparagraph]{Seán R. Kavanagh%
    \,\orcidlink{0000-0003-4577-9647}\,%
    }
\author[3%
  ]{Alexander G. Squires%
    \,\orcidlink{0000-0001-6967-3690}\,%
    }
\author[2%
  ]{Adair Nicolson%
    \,\orcidlink{0000-0002-8889-9369}\,%
    }
\author[1%
  ]{Irea Mosquera-Lois%
    \,\orcidlink{0000-0001-7651-0814}\,%
    }
\author[4%
  ]{Alex M. Ganose%
    \,\orcidlink{0000-0002-4486-3321}\,%
    }
\author[2%
  ]{Bonan Zhu%
    \,\orcidlink{0000-0001-5601-6130}\,%
    }
\author[2%
  ]{Katarina Brlec%
    \,\orcidlink{0000-0003-1485-1888}\,%
    }
\author[1%
  ]{Aron Walsh%
    \,\orcidlink{0000-0001-5460-7033}\,%
    }
\author[3%
  ]{David O. Scanlon%
    \,\orcidlink{0000-0001-9174-8601}\,%
    }

\affil[1]{Thomas Young Centre and Department of Materials, Imperial
College London, United Kingdom}
\affil[2]{Thomas Young Centre and Department of Chemistry, University
College London, United Kingdom}
\affil[3]{School of Chemistry, University of Birmingham, Birmingham,
United Kingdom}
\affil[4]{Department of Chemistry, Imperial College London, London,
United Kingdom}
\affil[$\mathparagraph$]{Corresponding author}
\date{01 March 2024}

\begin{document}
\maketitle

\hypertarget{summary}{%
\section{Summary}\label{summary}}

Defects are a universal feature of crystalline solids, dictating the key
properties and performance of many functional materials. Given their
crucial importance yet inherent difficulty in measuring experimentally,
computational methods (such as DFT and ML/classical force-fields) are
widely used to predict defect behaviour at the atomic level and the
resultant impact on macroscopic properties. Here we report
\texttt{doped}, a Python package for the generation,
pre-/post-processing, and analysis of defect supercell calculations.
\texttt{doped} has been built to implement the defect simulation
workflow in an efficient and user-friendly -- yet powerful and
fully-flexible -- manner, with the goal of providing a robust
general-purpose platform for conducting reproducible calculations of
solid-state defect properties.

\hypertarget{statement-of-need}{%
\section{Statement of need}\label{statement-of-need}}

The materials science sub-field of computational defect modelling has
seen considerable growth in recent years, driven by the crucial
importance of these species in functional materials and the major
advances in computational methodologies and resources facilitating their
accurate simulation. Software which enables researchers to efficiently
and accurately perform these calculations, while allowing for in-depth
target analyses of the resultant data, is thus of significant value to
the community. Indeed there are many critical stages in the
computational workflow for defects, which when performed manually not
only consume significant researcher time and effort but also leave room
for human error -- particularly for newcomers to the field. Moreover,
there are growing efforts to perform high-throughput investigations of
defects in solids
(\protect\hyperlink{ref-broberg_high-throughput_2023}{Broberg et al.,
2023}; \protect\hyperlink{ref-xiong_high-throughput_2023}{Xiong et al.,
2023}; \protect\hyperlink{ref-yuan_discovery_2024}{Yuan et al., 2024}),
necessitating robust, user-friendly, and efficient software implementing
this calculation workflow.

Given this importance of defect simulations and the complexity of the
workflow, a number of software packages have been developed with the
goal of managing pre- and post-processing of defect calculations,
including work on the \texttt{HADES}/\texttt{METADISE} codes from the
1970s (\protect\hyperlink{ref-parker_hades_2004}{Parker et al., 2004}),
to more recent work from Kumagai et al.
(\protect\hyperlink{ref-Kumagai2021}{2021}), Broberg et al.
(\protect\hyperlink{ref-Broberg2018}{2018}), Shen \& Varley
(\protect\hyperlink{ref-Shen2024}{2024}), Neilson \& Murphy
(\protect\hyperlink{ref-neilson_defap_2022}{2022}), Arrigoni \& Madsen
(\protect\hyperlink{ref-Arrigoni2021}{2021}), Goyal et al.
(\protect\hyperlink{ref-Goyal2017}{2017}), M. Huang et al.
(\protect\hyperlink{ref-Huang2022}{2022}), Péan et al.
(\protect\hyperlink{ref-pean_presentation_2017}{2017}) and Naik \& Jain
(\protect\hyperlink{ref-naik_coffee_2018}{2018}).\footnote{Some of these
  packages are no longer maintained, not compatible with high-throughput
  architectures, and/or are closed-source/commercial packages.} While
each of these codes have their strengths, they do not include the full
suite of functionality provided by \texttt{doped} -- some of which is
discussed below -- nor adopt the same focus on user-friendliness (along
with sanity-checking warnings and error catching) and efficiency with
full flexibility and wide-ranging functionality, targeting expert-level
users and newcomers to the field alike.

\begin{figure}
\centering
\includegraphics{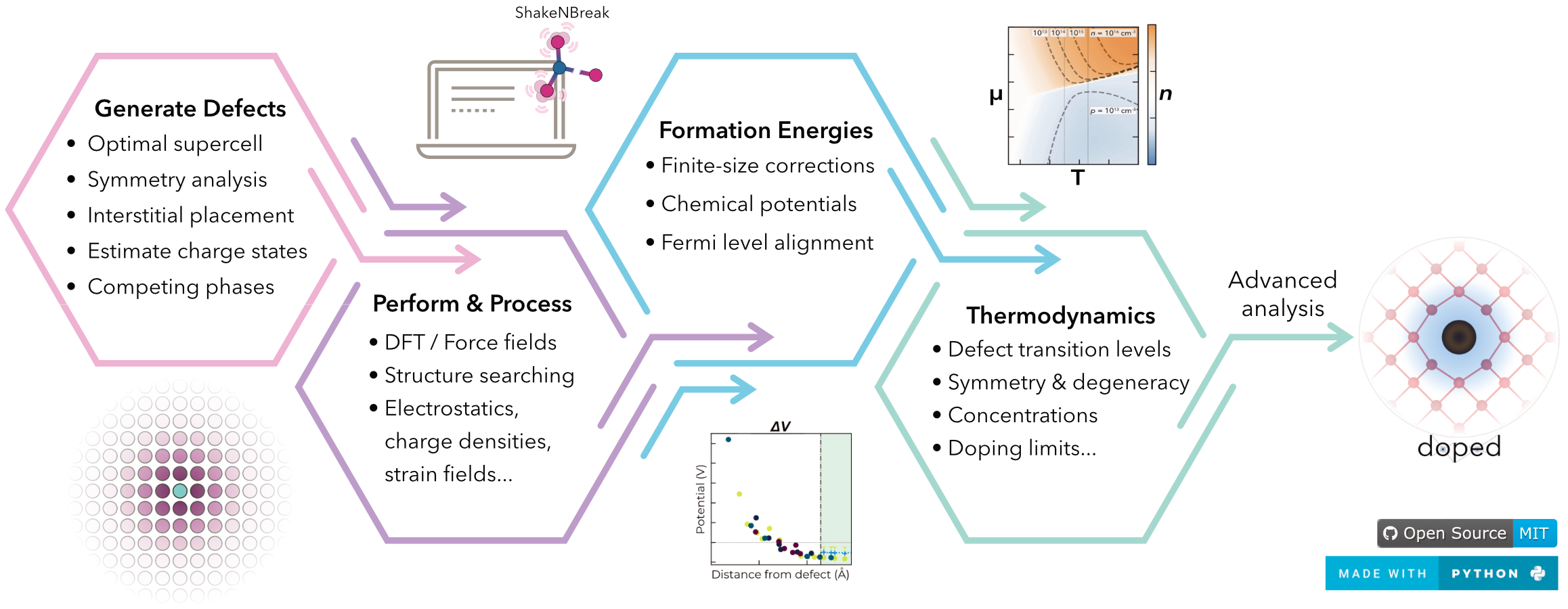}
\caption{Schematic workflow of a computational defect investigation
using \texttt{doped}. \label{fig_workflow}}
\end{figure}

\hypertarget{doped}{%
\section{doped}\label{doped}}

\texttt{doped} is a Python package for the generation,
pre-/post-processing, and analysis of defect supercell calculations, as
depicted in \autoref{fig_workflow}. The design philosophy of
\texttt{doped} has been to implement the defect simulation workflow in
an efficient, reproducible, and user-friendly -- yet powerful and
fully-customisable -- manner, combining reasonable defaults with full
user control for each parameter in the workflow. As depicted in
\autoref{fig_workflow}, the core functionality of \texttt{doped} is the
generation of defect supercells and competing phases, writing
calculation input files, parsing calculation outputs, and
analysing/plotting defect-related properties. This functionality and
recommended usage of \texttt{doped} is demonstrated in the
\href{https://doped.readthedocs.io/en/latest/Tutorials.html}{tutorials}
on the \href{https://doped.readthedocs.io/en/latest/}{documentation
website}.

\begin{figure}
\centering
\includegraphics{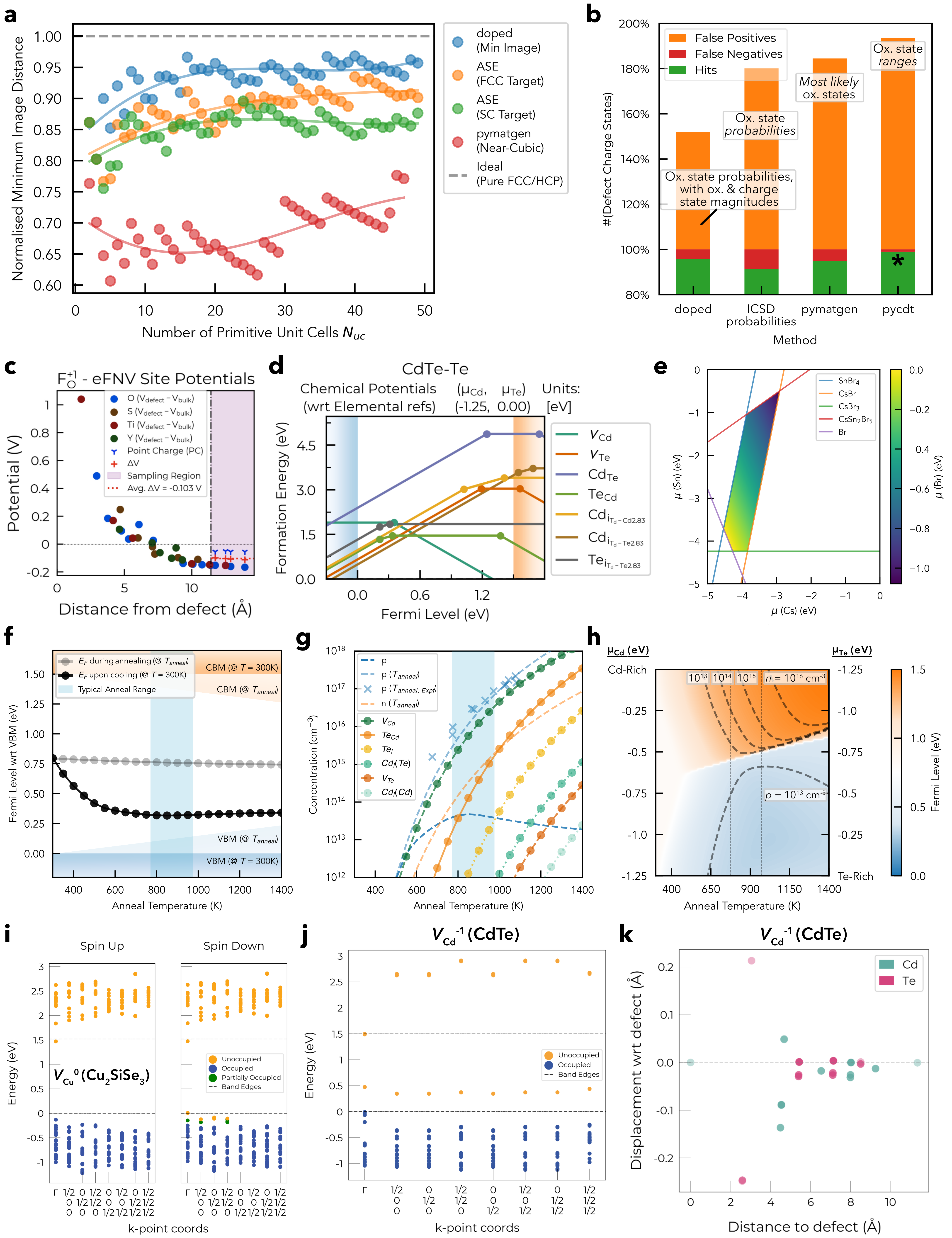}
\caption{\small{Performance and example outputs from \texttt{doped}.
\textbf{(a)} Average minimum periodic image distance, normalised by the
ideal image distance (i.e.~for a close-packed face-centred cubic (FCC)
cell), vs.~number of unit cells for supercell generation algorithms in
\texttt{doped}, \texttt{ASE}, and \texttt{pymatgen}. ``SC'' = simple
cubic and ``HCP'' = hexagonal close-packed. \textbf{(b)} Average
performance of various charge state estimation routines. ``ICSD
probabilities'' refers to a model based oxidation state probabilities,
as given by their occurrence in the ICSD database. Asterisk indicates
that \texttt{pyCDT} ``false \emph{negatives}'' are underestimated as the
majority of this test set used the \texttt{pyCDT} charge state ranges.
``Ox. state'' = oxidation state. Example \textbf{(c)} Kumagai-Oba (eFNV)
finite-size correction plot, \textbf{(d)} defect formation energy
diagram, \textbf{(e)} chemical potential / stability region,
\textbf{(f)} Fermi level vs.~annealing temperature, \textbf{(g)}
defect/carrier concentrations vs.~annealing temperature and \textbf{(h)}
Fermi level / carrier concentration heatmap plots from \texttt{doped}.
Automated plots of single-particle eigenvalues from DFT supercell
calculations for \textbf{(i)} \(V_{Cu}^{0}\) in Cu\(_2\)SiSe\(_3\) and
\textbf{(j)} \(V_{Cd}^{-1}\) in CdTe. \textbf{(k)} Automated site
displacement analysis, plotting atomic displacements with respect to the
defect site against distance to the defect site, for \(V_{Cd}^{-1}\) in
CdTe. Data and code to reproduce these plots is provided in the
\href{https://github.com/SMTG-Bham/doped/blob/main/docs/JOSS}{\texttt{docs/JOSS}}
folder of the \texttt{doped} GitHub repository. \label{fig1}}}
\end{figure}

Some key advances of \texttt{doped} include:

\begin{itemize}
\item
  \textbf{Supercell Generation:} When choosing a simulation supercell
  for charged defects in materials, we typically want to maximise the
  minimum distance between periodic images of the defect (to reduce
  finite-size errors) while keeping the supercell to a tractable number
  of atoms/electrons to calculate. Common approaches are to choose a
  near-cubic integer expansion of the unit cell
  (\protect\hyperlink{ref-ong_python_2013}{Ong et al., 2013}), or to use
  a cell shape metric to search for optimal supercells
  (\protect\hyperlink{ref-larsen_atomic_2017}{Larsen et al., 2017}).
  Building on these and instead integrating an efficient algorithm for
  calculating minimum image distances, \texttt{doped} directly optimises
  the supercell choice for this goal -- often identifying non-trivial
  `root 2'/`root 3' type supercells. As illustrated in \autoref{fig1}a,
  this leads to a significant reduction in the supercell size (and thus
  computational cost) required to achieve a threshold minimum image
  distance.

  \begin{itemize}
  \tightlist
  \item
    Over a test set of simple cubic, trigonal, orthorhombic, monoclinic
    and face-centred cubic unit cells, the \texttt{doped} algorithm is
    found to give mean improvements of 35.2\%, 9.1\% and 6.7\% in the
    minimum image distance for a given (maximum) number of unit cells as
    compared to the \texttt{pymatgen} cubic supercell algorithm, the
    \texttt{ASE} optimal cell shape algorithm with simple-cubic target
    shape, and \texttt{ASE} with FCC target shape respectively -- in the
    range of 2-20 unit cells. For 2-50 unit cells (for which the mean
    values across this test set are plotted in \autoref{fig1}a), this
    becomes 36.0\%, 9.3\% and 5.6\% respectively. Given the
    approximately cubic scaling of DFT computational cost with the
    number of atoms, these correspond to significant reductions in cost
    (\textasciitilde20-150\%).
  \item
    As always, the user has full control over supercell generation in
    \texttt{doped}, with the ability to specify/adjust constraints on
    the minimum image distance, number of atoms or transformation
    matrix, or to simply provide a pre-generated supercell if desired.
  \end{itemize}
\item
  \textbf{Charge-state Estimation:} Defects in solids can adopt various
  electronic charge states. However, the set of stable charge states for
  a given defect is typically not known \emph{a priori}, so one must
  choose a set of \emph{possible} defect charge states to calculate --
  usually relying on some form of chemical intuition. In this regard,
  extremal defect charge states that are calculated but do not end up
  being stable can be considered `false positives' or `wasted'
  calculations,\footnote{Note that \emph{unstable} defect charge states
    which are intermediate between \emph{stable} charge states
    (e.g.~\(X^0\) for a defect \(X\) with a (+1/-1) negative-U level)
    should still be calculated and are \emph{not} considered false
    positives.} while charge states which are stable but were not
  calculated can be considered `false negatives' or `missed'
  calculations. \texttt{doped} builds on other routines which use known
  elemental oxidation states to additionally account for oxidation state
  \emph{probabilities}, the electronic state of the host crystal and
  charge state magnitudes. Implementing these features in a simple cost
  function, we find a significant improvement in terms of both
  efficiency (reduced false positives) and completeness (reduced false
  negatives) for this charge state estimation, as shown in
  \autoref{fig1}b.\footnote{Given sufficient data, a machine learning
    model could likely further improve the performance of this charge
    state estimation.}

  Again, this step is fully-customisable. The user can tune the
  probability threshold at which to include charge states or manually
  specify defect charge states. All probability factors computed are
  available to the user and saved to the defect \texttt{JSON} files for
  full reproducibility.
\end{itemize}

\begin{itemize}
\item
  \textbf{Efficient Competing Phase Selection:} Elemental chemical
  potentials (a key term in the defect formation energy) are limited by
  the secondary phases which border the host compound on the phase
  diagram. These bordering phases are known as competing phases, and
  their total energies must be calculated to determine the chemical
  potential limits. Only the elemental reference phases and compounds
  which border the host on the phase diagram need to be calculated,
  rather than the full phase diagram.

  \texttt{doped} aims to improve the efficiency of this step by querying
  the \href{https://materialsproject.org}{Materials Project} database
  (containing both experimentally-measured and theoretically-predicted
  crystal structures), and pulling only compounds which \emph{could
  border the host material} within a user-specified error tolerance for
  the semi-local DFT database energies (0.1 eV/atom by default), along
  with the elemental reference phases. The necessary \emph{k}-point
  convergence step for these compounds is also implemented in a
  semi-automated fashion to expedite this process.

  \begin{itemize}
  \tightlist
  \item
    With the parsed chemical potentials in \texttt{doped}, the user can
    easily select various X-poor/rich chemical conditions, or scan over
    a range of chemical potentials (growth conditions) as shown in
    \autoref{fig1}e,h.
  \end{itemize}
\item
  \textbf{Automated Symmetry \& Degeneracy Handling:} \texttt{doped}
  automatically determines the point symmetry of both initial
  (un-relaxed) and final (relaxed) defect configurations, and computes
  the corresponding orientational (and spin) degeneracy factors. This
  functionality is also offered in the form of
  \href{https://doped.readthedocs.io/en/latest/advanced_analysis_tutorial.html\#point-symmetry-analysis}{standalone
  functions} which do not require the defect calculations to have been
  generated/parsed with \texttt{doped}. This is a key pre-factor in the
  defect concentration equation:

  \begin{equation}
  N_D = gN_s \exp(-E_f/k_BT)
  \end{equation}

  where \(g\) is the product of all degeneracy factors, \(N_s\) is the
  concentration of lattice sites for that defect, \(E_f\) is the defect
  formation energy and \(N_D\) is the defect concentration. \(g\) can
  affect predicted defect/carrier concentrations by up to two or three
  orders of magnitude
  (\protect\hyperlink{ref-kavanagh_impact_2022}{Kavanagh, Scanlon, et
  al., 2022};
  \protect\hyperlink{ref-mosquera-lois_imperfections_2023}{Mosquera-Lois,
  Kavanagh, Klarbring, et al., 2023}), and is often overlooked in defect
  calculations, partly due to the (previous) requirement of significant
  manual effort and knowledge of group theory.
\item
  \textbf{Automated Compatibility Checking:} When parsing defect
  calculations, \texttt{doped} automatically checks that calculation
  parameters which could affect the defect formation energy
  (e.g.~\emph{k}-point grid, energy cutoff, pseudopotential choice,
  exchange fraction, Hubbard U etc.) are consistent between the defect
  and reference calculations. This is a common source of accidental
  error in defect calculations, and \texttt{doped} provides informative
  warnings if any inconsistencies are detected.
\item
  \textbf{Thermodynamic Analysis:} \texttt{doped} provides a suite of
  flexible tools for the analysis of defect thermodynamics, including
  formation energy diagrams (\autoref{fig1}d), equilibrium \&
  non-equilibrium Fermi level solving (\autoref{fig1}f), doping analysis
  (\autoref{fig1}g,h), Brouwer-type diagrams etc. These include
  physically-motivated (but tunable) grouping of defect sites, full
  inclusion of metastable states, support for complex system
  constraints, optimisation over high-dimensional chemical \&
  temperature space and highly customisable plotting. In-depth examples
  are provided in the
  \href{https://doped.readthedocs.io/en/latest/Tutorials.html}{tutorials}.
\item
  \textbf{Finite-Size Corrections:} Both the isotropic Freysoldt (FNV)
  (\protect\hyperlink{ref-Freysoldt2009}{Freysoldt et al., 2009}) and
  anisotropic Kumagai (eFNV)
  (\protect\hyperlink{ref-kumagai_electrostatics-based_2014}{Kumagai \&
  Oba, 2014}) image charge corrections are implemented automatically in
  \texttt{doped}, with tunable sampling radii / sites (which may be
  desirable for e.g.~layered materials), automated correction plotting
  (to visualise/analyse convergence; \autoref{fig1}c), and automatic
  sampling error estimation.
\item
  \textbf{Reproducibility \& Tabulation:} \texttt{doped} has been built
  to support and encourage reproducibility, with all input parameters
  and calculation results saved to lightweight \texttt{JSON} files. This
  allows for easy sharing of calculation inputs/outputs and reproducible
  analysis. Several tabulation functions are also provided to facilitate
  the quick summarising of key quantities as exemplified in the
  \href{https://doped.readthedocs.io/en/latest/Tutorials.html}{tutorials}
  (including defect formation energy contributions, charge transition
  levels (with/without metastable states), symmetry, degeneracy and
  multiplicity factors, defect/carrier concentrations, chemical
  potential limits, dopability limits, doping windows\ldots) to aid
  transparency, reproducibility, comparisons with other works, and
  general analysis. The use of these tabulated outputs in supporting
  information of publications is encouraged.
\item
  \textbf{High-Throughput Compatibility:} \texttt{doped} is built to be
  compatible with high-throughput architectures such as
  \href{https://github.com/materialsproject/atomate2}{atomate(2)}
  (\protect\hyperlink{ref-atomate}{Mathew et al., 2017}) or
  \href{https://aiida.net}{AiiDA} (\protect\hyperlink{ref-AiiDA}{Huber
  et al., 2020}), aided by its object-oriented Python framework,
  JSON-serializable classes and sub-classed \texttt{pymatgen} objects.
  Examples are provided on the
  \href{https://doped.readthedocs.io/en/latest/}{documentation website}.
\item
  \textbf{\href{https://shakenbreak.readthedocs.io}{\texttt{ShakeNBreak}}:}
  \texttt{doped} is natively interfaced with our defect
  structure-searching code \texttt{ShakeNBreak}
  (\protect\hyperlink{ref-mosquera-lois_shakenbreak_2022}{Mosquera-Lois
  et al., 2022}), seamlessly incorporating this phase in the defect
  calculation workflow. This step can optionally be skipped or an
  alternative structure-searching approach readily implemented.
\end{itemize}

Some additional features of \texttt{doped} include directional-dependent
site displacement (local strain) analysis, deterministic \& informative
defect naming, molecule generation for gaseous competing phases,
multiprocessing for expedited generation \& parsing, shallow defect
analysis (via \texttt{pydefect}
(\protect\hyperlink{ref-Kumagai2021}{Kumagai et al., 2021})), Wyckoff
site analysis (including \emph{arbitrary/interstitial} sites),
controllable defect site placement to aid visualisation and more.

The defect generation and thermodynamic analysis components of
\texttt{doped} are agnostic to the underlying software used for the
defect supercell calculations. Direct calculation I/O is fully-supported
for \texttt{VASP} (\protect\hyperlink{ref-vasp}{Kresse \& Furthmüller,
1996}), while input defect structure files can be generated for several
widely-used DFT codes, including \texttt{FHI-aims}
(\protect\hyperlink{ref-fhi_aims}{Blum et al., 2009}), \texttt{CP2K}
(\protect\hyperlink{ref-cp2k}{Kühne et al., 2020}),
\texttt{Quantum\ Espresso} (\protect\hyperlink{ref-espresso}{Giannozzi
et al., 2009}) and \texttt{CASTEP} (\protect\hyperlink{ref-castep}{Clark
et al., 2005}) via the \texttt{pymatgen} \texttt{Structure} object. Full
support for calculation I/O with other DFT codes may be added in the
future if there is sufficient demand. Moreover, \texttt{doped} is built
to be readily compatible with other computational toolkits for advanced
defect characterisation, such as \texttt{ShakeNBreak} for defect
structure-searching, \texttt{py-sc-fermi} for advanced thermodynamic
analysis under complex constraints
(\protect\hyperlink{ref-squires_py-sc-fermi_2023}{Squires et al.,
2023}), \texttt{easyunfold} for analysing defect/dopant-induced
electronic structure changes
(\protect\hyperlink{ref-zhu_easyunfold_2024}{Zhu et al., 2024}) or
\texttt{CarrierCapture.jl}/\texttt{nonrad} for non-radiative
recombination calculations
(\protect\hyperlink{ref-kim_carriercapturejl_2020}{Kim et al., 2020};
\protect\hyperlink{ref-turiansky_nonrad_2021}{Turiansky et al., 2021}).

\texttt{doped} has been used to manage the defect simulation workflow in
a number of publications thus far, including Wang et al.
(\protect\hyperlink{ref-wang_upper_2024}{2024}), Cen et al.
(\protect\hyperlink{ref-cen_cation_2023}{2023}), Nicolson et al.
(\protect\hyperlink{ref-nicolson_cu2sise3_2023}{2023}), Li et al.
(\protect\hyperlink{ref-li_computational_2024}{2024}), Kumagai et al.
(\protect\hyperlink{ref-kumagai_alkali_2023}{2023}), Woo et al.
(\protect\hyperlink{ref-woo_inhomogeneous_2023}{2023}), Wang et al.
(\protect\hyperlink{ref-wang_four-electron_2023-1}{2023}), Mosquera-Lois
\& Kavanagh (\protect\hyperlink{ref-mosquera-lois_search_2021}{2021}),
Mosquera-Lois, Kavanagh, Walsh, et al.
(\protect\hyperlink{ref-mosquera-lois_identifying_2023}{2023}),
Mosquera-Lois et al.
(\protect\hyperlink{ref-mosquera-lois_machine-learning_2024}{2024}),
Y.-T. Huang et al. (\protect\hyperlink{ref-huang_strong_2022}{2022}),
Dou et al. (\protect\hyperlink{ref-dou_giant_2024}{2024}), Liga et al.
(\protect\hyperlink{ref-liga_mixed-cation_2023}{2023}), Willis, Spooner,
et al. (\protect\hyperlink{ref-willis_possibility_2023}{2023}), Willis,
Claes, et al. (\protect\hyperlink{ref-willis_limits_2023}{2023}),
Krajewska et al.
(\protect\hyperlink{ref-krajewska_enhanced_2021}{2021}), Kavanagh et al.
(\protect\hyperlink{ref-kavanagh_rapid_2021}{2021}), Kavanagh, Savory,
et al. (\protect\hyperlink{ref-kavanagh_frenkel_2022}{2022}).

\hypertarget{credit-author-contributions}{%
\section{CRediT Author
Contributions}\label{credit-author-contributions}}

\textbf{Seán R. Kavanagh:} Conceptualisation, Methodology, Software,
Writing, Project Administration. \textbf{Alex G. Squires:} Code for
complex doping analysis. \textbf{Adair Nicolson:} Code for shallow
defect analysis. \textbf{Irea Mosquera-Lois:} Code for local strain
analysis. \textbf{Katarina Brlec:} Competing phases code refactoring.
\textbf{Aron Walsh \& David Scanlon:} Funding Acquisition, Management,
Ideas \& Discussion. \textbf{All authors:} Feedback, Code Contributions,
Writing -- Review \& Editing.

\hypertarget{acknowledgements}{%
\section{Acknowledgements}\label{acknowledgements}}

\texttt{doped} has benefited from feature requests and feedback from
many members of the Walsh and Scanlon research groups, including (but
not limited to) Xinwei Wang, Sabrine Hachmioune, Savya Aggarwal, Daniel
Sykes, Chris Savory, Jiayi Cen, Lavan Ganeshkumar, Ke Li, Kieran Spooner
and Luisa Herring-Rodriguez. S.R.K thanks Dr.~Christoph Freysoldt and
Prof.~Yu Kumagai for useful discussions regarding the implementation of
image charge corrections.

The initial development of \texttt{doped} was inspired by the
\texttt{pyCDT} package from Broberg et al.
(\protect\hyperlink{ref-Broberg2018}{2018}), while the original colour
scheme for defect formation energy plots was inspired by work from Drs.
Adam J. Jackson and Alex M. Ganose. \texttt{doped} makes extensive use
of Python objects from the widely-used \texttt{pymatgen}
(\protect\hyperlink{ref-ong_python_2013}{Ong et al., 2013}) package
(such as structure representations and VASP I/O handling), as well as
crystal symmetry functions from \texttt{spglib}
(\protect\hyperlink{ref-togo_textttspglib_2018}{Togo \& Tanaka, 2018}).

S.R.K. and A.N. acknowledge the EPSRC Centre for Doctoral Training in
the Advanced Characterisation of Materials (CDTACM)(EP/S023259/1) for
funding PhD studentships. DOS acknowledges support from the EPSRC
(EP/N01572X/1) and from the European Research Council, ERC (Grant
No.~758345). The PRAETORIAN project was funded by UK Research and
Innovation (UKRI) under the UK government's Horizon Europe funding
guarantee (EP/Y019504/1). This work used the ARCHER2 UK National
Supercomputing Service (https://www.archer2.ac.uk), via our membership
of the UK's HEC Materials Chemistry Consortium, which is funded by the
EPSRC (EP/L000202, EP/R029431 and EP/T022213), the UK Materials and
Molecular Modelling (MMM) Hub (Young EP/T022213).

\hypertarget{references}{%
\section*{References}\label{references}}
\addcontentsline{toc}{section}{References}

\hypertarget{refs}{}
\begin{CSLReferences}{1}{0}
\leavevmode\vadjust pre{\hypertarget{ref-Arrigoni2021}{}}%
Arrigoni, M., \& Madsen, G. K. H. (2021). {Spinney: Post-processing of
first-principles calculations of point defects in semiconductors with
Python}. \emph{Computer Physics Communications}, \emph{264}, 107946.
\url{https://doi.org/10.1016/j.cpc.2021.107946}

\leavevmode\vadjust pre{\hypertarget{ref-fhi_aims}{}}%
Blum, V., Gehrke, R., Hanke, F., Havu, P., Havu, V., Ren, X., Reuter,
K., \& Scheffler, M. (2009). Ab initio molecular simulations with
numeric atom-centered orbitals. \emph{Computer Physics Communications},
\emph{180}(11), 2175--2196.
\url{https://doi.org/10.1016/j.cpc.2009.06.022}

\leavevmode\vadjust pre{\hypertarget{ref-broberg_high-throughput_2023}{}}%
Broberg, D., Bystrom, K., Srivastava, S., Dahliah, D., Williamson, B. A.
D., Weston, L., Scanlon, D. O., Rignanese, G.-M., Dwaraknath, S.,
Varley, J., Persson, K. A., Asta, M., \& Hautier, G. (2023).
High-throughput calculations of charged point defect properties with
semi-local density functional theory -- performance benchmarks for
materials screening applications. \emph{Npj Computational Materials},
\emph{9}(1), 1--12. \url{https://doi.org/10.1038/s41524-023-01015-6}

\leavevmode\vadjust pre{\hypertarget{ref-Broberg2018}{}}%
Broberg, D., Medasani, B., Zimmermann, N. E. R., Yu, G., Canning, A.,
Haranczyk, M., Asta, M., \& Hautier, G. (2018). PyCDT: A python toolkit
for modeling point defects in semiconductors and insulators.
\emph{Computer Physics Communications}, \emph{226}, 165--179.
\url{https://doi.org/10.1016/j.cpc.2018.01.004}

\leavevmode\vadjust pre{\hypertarget{ref-cen_cation_2023}{}}%
Cen, J., Zhu, B., R. Kavanagh, S., G. Squires, A., \& O. Scanlon, D.
(2023). Cation disorder dominates the defect chemistry of high-voltage
{LiMn}{\textsubscript{1.5}}{Ni}{\textsubscript{0.5}}{O}{\textsubscript{4}}({LMNO})
spinel cathodes. \emph{Journal of Materials Chemistry A}, \emph{11}(25),
13353--13370. \url{https://doi.org/10.1039/D3TA00532A}

\leavevmode\vadjust pre{\hypertarget{ref-castep}{}}%
Clark, S. J., Segall, M. D., Pickard, C. J., Hasnip, P. J., P., M. I.
J., Refson, K., \& Payne, M. C. (2005). {First principles methods using
CASTEP}. \emph{Zeitschrift Für Kristallographie - Crystalline
Materials}, \emph{220}(5-6), 567--570.
\url{https://doi.org/10.1524/zkri.220.5.567.65075}

\leavevmode\vadjust pre{\hypertarget{ref-dou_giant_2024}{}}%
Dou, W., Spooner, K., Kavanagh, S., Zhou, M., \& Scanlon, D. O. (2024).
Giant {Band Degeneracy} via {Orbital Engineering Enhances Thermoelectric
Performance} from
{Sb{\textsubscript{2}}Si{\textsubscript{2}}Te{\textsubscript{6}}} to
{Sc{\textsubscript{2}}Si{\textsubscript{2}}Te{\textsubscript{6}}}.
\emph{{ChemRxiv}}. \url{https://doi.org/10.26434/chemrxiv-2024-hm6vh}

\leavevmode\vadjust pre{\hypertarget{ref-Freysoldt2009}{}}%
Freysoldt, C., Neugebauer, J., \& Walle, C. V. de. (2009). Fully ab
initio finite-size corrections for charged-defect supercell
calculations. \emph{Physical Review Letters}, \emph{102}, 016402.
\url{https://doi.org/10.1103/PhysRevLett.102.016402}

\leavevmode\vadjust pre{\hypertarget{ref-espresso}{}}%
Giannozzi, P., Baroni, S., Bonini, N., Calandra, M., Car, R., Cavazzoni,
C., Ceresoli, D., Chiarotti, G. L., Cococcioni, M., Dabo, I., Corso, A.
D., Gironcoli, S. de, Fabris, S., Fratesi, G., Gebauer, R., Gerstmann,
U., Gougoussis, C., Kokalj, A., Lazzeri, M., \ldots{} Wentzcovitch, R.
M. (2009). {QUANTUM ESPRESSO: a modular and open-source software project
for quantum simulations of materials}. \emph{Journal of Physics:
Condensed Matter}, \emph{21}(39), 395502.
\url{https://doi.org/10.1088/0953-8984/21/39/395502}

\leavevmode\vadjust pre{\hypertarget{ref-Goyal2017}{}}%
Goyal, A., Gorai, P., Peng, H., Lany, S., \& Stevanović, V. (2017). {A
computational framework for automation of point defect calculations}.
\emph{Computational Materials Science}, \emph{130}, 1--9.
\url{https://doi.org/10.1016/j.commatsci.2016.12.040}

\leavevmode\vadjust pre{\hypertarget{ref-Huang2022}{}}%
Huang, M., Zheng, Z., Dai, Z., Guo, X., Wang, S., Jiang, L., Wei, J., \&
Chen, S. (2022). DASP: Defect and dopant ab-initio simulation package.
\emph{Journal of Semiconductors}, \emph{43}, 42101.
\url{https://doi.org/10.1088/1674-4926/43/4/042101}

\leavevmode\vadjust pre{\hypertarget{ref-huang_strong_2022}{}}%
Huang, Y.-T., Kavanagh, S. R., Righetto, M., Rusu, M., Levine, I.,
Unold, T., Zelewski, S. J., Sneyd, A. J., Zhang, K., Dai, L., Britton,
A. J., Ye, J., Julin, J., Napari, M., Zhang, Z., Xiao, J., Laitinen, M.,
Torrente-Murciano, L., Stranks, S. D., \ldots{} Hoye, R. L. Z. (2022).
Strong absorption and ultrafast localisation in
{NaBiS}{\textsubscript{2}} nanocrystals with slow charge-carrier
recombination. \emph{Nature Communications}, \emph{13}(1), 4960.
\url{https://doi.org/10.1038/s41467-022-32669-3}

\leavevmode\vadjust pre{\hypertarget{ref-AiiDA}{}}%
Huber, S. P., Zoupanos, S., Uhrin, M., Talirz, L., Kahle, L.,
Häuselmann, R., Gresch, D., Müller, T., Yakutovich, A. V., Andersen, C.
W., Ramirez, F. F., Adorf, C. S., Gargiulo, F., Kumbhar, S., Passaro,
E., Johnston, C., Merkys, A., Cepellotti, A., Mounet, N., \ldots{}
Pizzi, G. (2020). {AiiDA 1.0, a scalable computational infrastructure
for automated reproducible workflows and data provenance}.
\emph{Scientific Data}, \emph{7}(300), 1--18.
\url{https://doi.org/10.1038/s41597-020-00638-4}

\leavevmode\vadjust pre{\hypertarget{ref-kavanagh_frenkel_2022}{}}%
Kavanagh, S. R., Savory, C. N., Liga, S. M., Konstantatos, G., Walsh,
A., \& Scanlon, D. O. (2022). Frenkel {Excitons} in {Vacancy-Ordered
Titanium Halide Perovskites}
({Cs{\textsubscript{2}}TiX{\textsubscript{6}}}). \emph{The Journal of
Physical Chemistry Letters}, \emph{13}(47), 10965--10975.
\url{https://doi.org/10.1021/acs.jpclett.2c02436}

\leavevmode\vadjust pre{\hypertarget{ref-kavanagh_impact_2022}{}}%
Kavanagh, S. R., Scanlon, D. O., Walsh, A., \& Freysoldt, C. (2022).
Impact of metastable defect structures on carrier recombination in solar
cells. \emph{Faraday Discussions}, \emph{239}(0), 339--356.
\url{https://doi.org/10.1039/D2FD00043A}

\leavevmode\vadjust pre{\hypertarget{ref-kavanagh_rapid_2021}{}}%
Kavanagh, S. R., Walsh, A., \& Scanlon, D. O. (2021). Rapid
{Recombination} by {Cadmium Vacancies} in {CdTe}. \emph{ACS Energy
Letters}, \emph{6}(4), 1392--1398.
\url{https://doi.org/10.1021/acsenergylett.1c00380}

\leavevmode\vadjust pre{\hypertarget{ref-kim_carriercapturejl_2020}{}}%
Kim, S., Hood, S. N., Gerwen, P. van, Whalley, L. D., \& Walsh, A.
(2020). {CarrierCapture}.jl: {Anharmonic Carrier Capture}. \emph{Journal
of Open Source Software}, \emph{5}(47), 2102.
\url{https://doi.org/10.21105/joss.02102}

\leavevmode\vadjust pre{\hypertarget{ref-krajewska_enhanced_2021}{}}%
Krajewska, C. J., Kavanagh, S. R., Zhang, L., Kubicki, D. J., Dey, K.,
Gałkowski, K., Grey, C. P., Stranks, S. D., Walsh, A., Scanlon, D. O.,
\& Palgrave, R. G. (2021). Enhanced visible light absorption in layered
{Cs}{\textsubscript{3}}{Bi}{\textsubscript{2}}{Br}{\textsubscript{9}}
through mixed-valence {Sn}({\textsc{II}})/{Sn}({\textsc{IV}}) doping.
\emph{Chemical Science}, \emph{12}(44), 14686--14699.
\url{https://doi.org/10.1039/D1SC03775G}

\leavevmode\vadjust pre{\hypertarget{ref-vasp}{}}%
Kresse, G., \& Furthmüller, J. (1996). {Efficient iterative schemes for
ab initio total-energy calculations using a plane-wave basis set}.
\emph{Physical Review B}, \emph{54}(16), 11169.
\url{https://doi.org/10.1103/PhysRevB.54.11169}

\leavevmode\vadjust pre{\hypertarget{ref-cp2k}{}}%
Kühne, T. D., Iannuzzi, M., Del Ben, M., Rybkin, V. V., Seewald, P.,
Stein, F., Laino, T., Khaliullin, R. Z., Schütt, O., Schiffmann, F.,
Golze, D., Wilhelm, J., Chulkov, S., Bani-Hashemian, M. H., Weber, V.,
Borštnik, U., Taillefumier, M., Jakobovits, A. S., Lazzaro, A., \ldots{}
Hutter, J. (2020). {CP2K: An electronic structure and molecular dynamics
software package - Quickstep: Efficient and accurate electronic
structure calculations}. \emph{The Journal of Chemical Physics},
\emph{152}(19), 194103. \url{https://doi.org/10.1063/5.0007045}

\leavevmode\vadjust pre{\hypertarget{ref-kumagai_alkali_2023}{}}%
Kumagai, Y., Kavanagh, S. R., Suzuki, I., Omata, T., Walsh, A., Scanlon,
D. O., \& Morito, H. (2023). Alkali {Mono-Pnictides}: {A New Class} of
{Photovoltaic Materials} by {Element Mutation}. \emph{PRX Energy},
\emph{2}(4), 043002. \url{https://doi.org/10.1103/PRXEnergy.2.043002}

\leavevmode\vadjust pre{\hypertarget{ref-kumagai_electrostatics-based_2014}{}}%
Kumagai, Y., \& Oba, F. (2014). Electrostatics-based finite-size
corrections for first-principles point defect calculations.
\emph{Physical Review B}, \emph{89}(19), 195205.
\url{https://doi.org/10.1103/PhysRevB.89.195205}

\leavevmode\vadjust pre{\hypertarget{ref-Kumagai2021}{}}%
Kumagai, Y., Tsunoda, N., Takahashi, A., \& Oba, F. (2021). Insights
into oxygen vacancies from high-throughput first-principles
calculations. \emph{Physical Review Materials}, \emph{5}, 123803.
\url{https://doi.org/10.1103/PhysRevMaterials.5.123803}

\leavevmode\vadjust pre{\hypertarget{ref-larsen_atomic_2017}{}}%
Larsen, A. H., Mortensen, J. J., Blomqvist, J., Castelli, I. E.,
Christensen, R., Du\textbackslash lak, M., Friis, J., Groves, M. N.,
Hammer, B., Hargus, C., Hermes, E. D., Jennings, P. C., Jensen, P. B.,
Kermode, J., Kitchin, J. R., Kolsbjerg, E. L., Kubal, J., Kaasbjerg, K.,
Lysgaard, S., \ldots{} Jacobsen, K. W. (2017). The atomic simulation
environment{\textemdash}a {Python} library for working with atoms.
\emph{Journal of Physics: Condensed Matter}, \emph{29}(27), 273002.
\url{https://doi.org/10.1088/1361-648X/aa680e}

\leavevmode\vadjust pre{\hypertarget{ref-li_computational_2024}{}}%
Li, K., Willis, J., Kavanagh, S. R., \& Scanlon, D. O. (2024).
Computational {Prediction} of an {Antimony-Based} n-{Type Transparent
Conducting Oxide}: {F-Doped Sb{\textsubscript{2}}O{\textsubscript{5}}}.
\emph{Chemistry of Materials}, \emph{36}(6), 2907--2916.
\url{https://doi.org/10.1021/acs.chemmater.3c03257}

\leavevmode\vadjust pre{\hypertarget{ref-liga_mixed-cation_2023}{}}%
Liga, S. M., Kavanagh, S. R., Walsh, A., Scanlon, D. O., \&
Konstantatos, G. (2023). Mixed-{Cation Vacancy-Ordered Perovskites}
({Cs{\textsubscript{2}}Ti{\textsubscript{1-x}}Sn{\textsubscript{x}}X{\textsubscript{6}}};
{X} = {I} or {Br}): {Low-Temperature Miscibility}, {Additivity}, and
{Tunable Stability}. \emph{The Journal of Physical Chemistry C},
\emph{127}(43), 21399--21409.
\url{https://doi.org/10.1021/acs.jpcc.3c05204}

\leavevmode\vadjust pre{\hypertarget{ref-atomate}{}}%
Mathew, K., Montoya, J. H., Faghaninia, A., Dwarakanath, S., Aykol, M.,
Tang, H., Chu, I., Smidt, T., Bocklund, B., Horton, M., Dagdelen, J.,
Wood, B., Liu, Z.-K., Neaton, J., Ong, S. P., Persson, K., \& Jain, A.
(2017). Atomate: A high-level interface to generate, execute, and
analyze computational materials science workflows. \emph{Computational
Materials Science}, \emph{139}, 140--152.
\url{https://doi.org/10.1016/j.commatsci.2017.07.030}

\leavevmode\vadjust pre{\hypertarget{ref-mosquera-lois_search_2021}{}}%
Mosquera-Lois, I., \& Kavanagh, S. R. (2021). In search of hidden
defects. \emph{Matter}, \emph{4}(8), 2602--2605.
\url{https://doi.org/10.1016/j.matt.2021.06.003}

\leavevmode\vadjust pre{\hypertarget{ref-mosquera-lois_machine-learning_2024}{}}%
Mosquera-Lois, I., Kavanagh, S. R., Ganose, A. M., \& Walsh, A. (2024).
Machine-learning structural reconstructions for accelerated point defect
calculations. \emph{{arXiv}}, \emph{arXiv:2401.12127}.
\url{https://doi.org/10.48550/arXiv.2401.12127}

\leavevmode\vadjust pre{\hypertarget{ref-mosquera-lois_imperfections_2023}{}}%
Mosquera-Lois, I., Kavanagh, S. R., Klarbring, J., Tolborg, K., \&
Walsh, A. (2023). Imperfections are not 0 {K}: Free energy of point
defects in crystals. \emph{Chemical Society Reviews}, \emph{52}(17),
5812--5826. \url{https://doi.org/10.1039/D3CS00432E}

\leavevmode\vadjust pre{\hypertarget{ref-mosquera-lois_shakenbreak_2022}{}}%
Mosquera-Lois, I., Kavanagh, S. R., Walsh, A., \& Scanlon, D. O. (2022).
{ShakeNBreak}: {Navigating} the defect configurational landscape.
\emph{Journal of Open Source Software}, \emph{7}(80), 4817.
\url{https://doi.org/10.21105/joss.04817}

\leavevmode\vadjust pre{\hypertarget{ref-mosquera-lois_identifying_2023}{}}%
Mosquera-Lois, I., Kavanagh, S. R., Walsh, A., \& Scanlon, D. O. (2023).
Identifying the ground state structures of point defects in solids.
\emph{Npj Computational Materials}, \emph{9}(1), 1--11.
\url{https://doi.org/10.1038/s41524-023-00973-1}

\leavevmode\vadjust pre{\hypertarget{ref-naik_coffee_2018}{}}%
Naik, M. H., \& Jain, M. (2018). {CoFFEE}: {Corrections For Formation
Energy} and {Eigenvalues} for charged defect simulations. \emph{Computer
Physics Communications}, \emph{226}, 114--126.
\url{https://doi.org/10.1016/j.cpc.2018.01.011}

\leavevmode\vadjust pre{\hypertarget{ref-neilson_defap_2022}{}}%
Neilson, W. D., \& Murphy, S. T. (2022). {DefAP}: {A Python} code for
the analysis of point defects in crystalline solids. \emph{Computational
Materials Science}, \emph{210}, 111434.
\url{https://doi.org/10.1016/j.commatsci.2022.111434}

\leavevmode\vadjust pre{\hypertarget{ref-nicolson_cu2sise3_2023}{}}%
Nicolson, A., Kavanagh, S. R., Savory, C. N., Watson, G. W., \& Scanlon,
D. O. (2023). {Cu{\textsubscript{2}}SiSe{\textsubscript{3}}} as a
promising solar absorber: Harnessing cation dissimilarity to avoid
killer antisites. \emph{Journal of Materials Chemistry A},
\emph{11}(27), 14833--14839. \url{https://doi.org/10.1039/D3TA02429F}

\leavevmode\vadjust pre{\hypertarget{ref-ong_python_2013}{}}%
Ong, S. P., Richards, W. D., Jain, A., Hautier, G., Kocher, M., Cholia,
S., Gunter, D., Chevrier, V. L., Persson, K. A., \& Ceder, G. (2013).
Python {Materials Genomics} (pymatgen): {A} robust, open-source python
library for materials analysis. \emph{Computational Materials Science},
\emph{68}, 314--319.
\url{https://doi.org/10.1016/j.commatsci.2012.10.028}

\leavevmode\vadjust pre{\hypertarget{ref-parker_hades_2004}{}}%
Parker, S. C., Cooke, D. J., Kerisit, S., Marmier, A. S., Taylor, S. L.,
\& Taylor, S. N. (2004). From {HADES} to
{PARADISE}{\textemdash}atomistic simulation of defects in minerals.
\emph{Journal of Physics: Condensed Matter}, \emph{16}(27), S2735.
\url{https://doi.org/10.1088/0953-8984/16/27/010}

\leavevmode\vadjust pre{\hypertarget{ref-pean_presentation_2017}{}}%
Péan, E., Vidal, J., Jobic, S., \& Latouche, C. (2017). Presentation of
the {PyDEF} post-treatment {Python} software to compute publishable
charts for defect energy formation. \emph{Chemical Physics Letters},
\emph{671}, 124--130. \url{https://doi.org/10.1016/j.cplett.2017.01.001}

\leavevmode\vadjust pre{\hypertarget{ref-Shen2024}{}}%
Shen, J.-X., \& Varley, J. (2024). Pymatgen-analysis-defects: {A Python}
package for analyzing point defects in crystalline materials.
\emph{Journal of Open Source Software}, \emph{9}(93), 5941.
\url{https://doi.org/10.21105/joss.05941}

\leavevmode\vadjust pre{\hypertarget{ref-squires_py-sc-fermi_2023}{}}%
Squires, A. G., Scanlon, D. O., \& Morgan, B. J. (2023). Py-sc-fermi:
Self-consistent {Fermi} energies and defect concentrations from
electronic structure calculations. \emph{Journal of Open Source
Software}, \emph{8}(82), 4962. \url{https://doi.org/10.21105/joss.04962}

\leavevmode\vadjust pre{\hypertarget{ref-togo_textttspglib_2018}{}}%
Togo, A., \& Tanaka, I. (2018). {\texttt{spglib}}: A software library
for crystal symmetry search. \emph{{arXiv}}, \emph{arXiv:1808.01590}.
\url{https://doi.org/10.48550/arXiv.1808.01590}

\leavevmode\vadjust pre{\hypertarget{ref-turiansky_nonrad_2021}{}}%
Turiansky, M. E., Alkauskas, A., Engel, M., Kresse, G., Wickramaratne,
D., Shen, J.-X., Dreyer, C. E., \& Van de Walle, C. G. (2021). Nonrad:
{Computing} nonradiative capture coefficients from first principles.
\emph{Computer Physics Communications}, \emph{267}, 108056.
\url{https://doi.org/10.1016/j.cpc.2021.108056}

\leavevmode\vadjust pre{\hypertarget{ref-wang_four-electron_2023-1}{}}%
Wang, X., Kavanagh, S. R., Scanlon, D. O., \& Walsh, A. (2023).
Four-electron negative-{\emph{U}} vacancy defects in antimony selenide.
\emph{Physical Review B}, \emph{108}(13), 134102.
\url{https://doi.org/10.1103/PhysRevB.108.134102}

\leavevmode\vadjust pre{\hypertarget{ref-wang_upper_2024}{}}%
Wang, X., Kavanagh, S. R., Scanlon, D. O., \& Walsh, A. (2024). Upper
efficiency limit of {Sb{\textsubscript{2}}Se{\textsubscript{3}}} solar
cells. \emph{{arXiv}}, \emph{arXiv:2402.04434}.
\url{https://doi.org/10.48550/arXiv.2402.04434}

\leavevmode\vadjust pre{\hypertarget{ref-willis_limits_2023}{}}%
Willis, J., Claes, R., Zhou, Q., Giantomassi, M., Rignanese, G.-M.,
Hautier, G., \& Scanlon, D. O. (2023). Limits to {Hole Mobility} and
{Doping} in {Copper Iodide}. \emph{Chemistry of Materials},
\emph{35}(21), 8995--9006.
\url{https://doi.org/10.1021/acs.chemmater.3c01628}

\leavevmode\vadjust pre{\hypertarget{ref-willis_possibility_2023}{}}%
Willis, J., Spooner, K. B., \& Scanlon, D. O. (2023). On the possibility
of p-type doping in barium stannate. \emph{Applied Physics Letters},
\emph{123}(16), 162103. \url{https://doi.org/10.1063/5.0170552}

\leavevmode\vadjust pre{\hypertarget{ref-woo_inhomogeneous_2023}{}}%
Woo, Y. W., Li, Z., Jung, Y.-K., Park, J.-S., \& Walsh, A. (2023).
Inhomogeneous {Defect Distribution} in {Mixed-Polytype Metal Halide
Perovskites}. \emph{ACS Energy Letters}, \emph{8}(1), 356--360.
\url{https://doi.org/10.1021/acsenergylett.2c02306}

\leavevmode\vadjust pre{\hypertarget{ref-xiong_high-throughput_2023}{}}%
Xiong, Y., Bourgois, C., Sheremetyeva, N., Chen, W., Dahliah, D., Song,
H., Zheng, J., Griffin, S. M., Sipahigil, A., \& Hautier, G. (2023).
High-throughput identification of spin-photon interfaces in silicon.
\emph{Science Advances}, \emph{9}(40), eadh8617.
\url{https://doi.org/10.1126/sciadv.adh8617}

\leavevmode\vadjust pre{\hypertarget{ref-yuan_discovery_2024}{}}%
Yuan, Z., Dahliah, D., Hasan, M. R., Kassa, G., Pike, A., Quadir, S.,
Claes, R., Chandler, C., Xiong, Y., Kyveryga, V., Yox, P., Rignanese,
G.-M., Dabo, I., Zakutayev, A., Fenning, D. P., Reid, O. G., Bauers, S.,
Liu, J., Kovnir, K., \& Hautier, G. (2024). Discovery of the
{Zintl-phosphide BaCd{\textsubscript{2}}P{\textsubscript{2}}} as a long
carrier lifetime and stable solar absorber. \emph{Joule}.
\url{https://doi.org/10.1016/j.joule.2024.02.017}

\leavevmode\vadjust pre{\hypertarget{ref-zhu_easyunfold_2024}{}}%
Zhu, B., Kavanagh, S. R., \& Scanlon, D. (2024). Easyunfold: {A Python}
package for unfolding electronic band structures. \emph{Journal of Open
Source Software}, \emph{9}(93), 5974.
\url{https://doi.org/10.21105/joss.05974}

\end{CSLReferences}

\end{document}